\documentclass[pra, twocolumn,showpacs,preprintnumbers,amsmath,amssymb]{revtex4}
\usepackage{dcolumn}
\usepackage{bm}
\usepackage{graphicx,subfigure,xcolor}

\usepackage{braket}
\usepackage{xcolor}

\usepackage{color, soul}
\setstcolor{red}
\setstcolor{blue}

\def\tp{\tau'}
\def\xt{x(\tau)}
\def\xtp{x(\tau')}
\def\xAt{x_A(\tau)}
\def\xAtp{x_A(\tau')}

\def\xBtp{x_B(\tau')}
\def\w{\omega}
\def\w0{\omega_0}
\def\wk{\omega_k}
\def\bk{\bf k}
\def\bx{\bf x}
\def\ak{a_{\bk}}
\def\akd{a_{\bk}^{\dagger}}

\def\g{\mid g \rangle}

\def\e{\mid e \rangle}

\def\f{\mbox{\small{\em f}}}

\def\vac{\mid 0\rangle}
\def\vacd{\langle 0\mid}

\begin{document}

\title{Non-thermal effects of acceleration in the resonance interaction between two uniformly accelerated atoms}
\author{Lucia Rizzuto$^{1,2}$, Margherita Lattuca$^{1}$, Jamir Marino$^{3,4}$, Antonio Noto$^{1,5}$,\\
Salvatore Spagnolo$^{1}$, Wenting Zhou$^{1,6}$, and Roberto Passante$^{1,2}$}
\address{$^1$ Dipartimento di Fisica e Chimica, Universit\`{a} degli Studi di Palermo and CNISM, Via Archirafi 36, I-90123 Palermo, Italy}
\email{lucia.rizzuto@unipa.it}
\address{$^2$ INFN, Laboratori Nazionali del Sud, I-95123 Catania, Italy}
\address{$^3$ Institute of Theoretical Physics, TU Dresden, D-01062 Dresden, Germany}
\address{$^4$ Institute of Theoretical Physics, University of Cologne, D-50937 Cologne, Germany}
\address{$^5$ Laboratoire Charles Coulomb UMR 5221 CNRS-UM2, D\'{e}partement Physique Th\'{e}orique, Universit\'{e} Montpellier 2, F-34095, Montpellier Cedex 5, France}
\address{$^6$ Center for Nonlinear Science and Department of Physics, Ningbo University, Ningbo, Zhejiang, 315211, China}


\begin{abstract}
We study the resonance interaction between two uniformly accelerated identical atoms, one excited and the other in the ground state, prepared in a correlated (symmetric or antisymmetric) state and interacting with the scalar field or the electromagnetic field in the vacuum state. In this case (resonance interaction), the interatomic interaction is a second-order effect in the atom-field coupling.
We separate the contributions of vacuum fluctuations and radiation reaction to the resonance energy shift of the system, and show that only radiation reaction contributes, while
Unruh thermal fluctuations do not affect the resonance interaction.
We also find that beyond a characteristic length scale related to the atomic acceleration, non-thermal-like effects in the radiation reaction contribution change the distance-dependence of the resonance interaction.
Finally, we find that previously unidentified features appear, compared with the scalar field case, when the interaction with the electromagnetic field is considered,
as a consequence of the peculiar nature of the vacuum quantum noise of the electromagnetic field in a relativistically accelerated background.
\end{abstract}

\maketitle

\section{\label{sec:1}Introduction}
A striking consequence of quantum field theory in non-Minkowskian spacetimes is the observer-dependent particle content of the quantum vacuum. An archetypical manifestation of this feature is the Unruh effect: a uniformly accelerated observer in the Minkowski vacuum perceives vacuum fluctuations as a thermal field, with a temperature $T_U$ proportional to its acceleration, $T_U=\frac{\hbar a}{2\pi k_Bc}$ \cite{Unruh76, Fulling73, CHM08}. The Unruh effect has stimulated  intense theoretical  investigations in last 40 years, including connections with cosmology \cite{Hawking74, Davies01} and applications to quantum optics and quantum information \cite{MA03,FSM05}.
Experimental proposals for its measurement using particles moving in circular accelerators \cite{NHLMM14}, or electrons accelerated by ultrastrong laser fields \cite{SSH06}, as well as in analog models of condensed matter physics \cite{RCPR08}, have also been suggested. However, despite these intense efforts the problem of detecting the Unruh effect remains open because of the very high accelerations, of the order of $\sim 10^{20}m/s^2$, necessary to obtain a Unruh temperature of  a few Kelvins. On the other hand, a direct verification of the Unruh effect could allow a deeper understanding of some controversies about its interpretation \cite{VM01, NFKMB01, FC06}.
In this direction, it has been recently argued that interatomic van der Waals/Casimir-Polder interactions between two uniformly accelerated atoms could
be very promising candidates for detecting the Unruh effect, even with reasonable values of the acceleration \cite{NP13,MNPRS14}.

Among the large number of open fundamental challenges in this field, a long-standing question is whether or not the effect of a relativistic acceleration is strictly equivalent to a thermal field \cite{BS13,MNP14}. It has been recently shown, for example, that non-thermal features associated with uniform acceleration manifest in the radiative properties of single accelerated atoms \cite{YZ06, Passante98, Rizzuto07, ZY10, MNPRS14, RS09}.
Also, recent works on entanglement generation or Casimir-Polder interactions between uniformly accelerated atoms, have shown that non-thermal effects of acceleration  arise in a system of two or many particles  \cite{HY15, BF04, MNP14}.

In this context, it has been recently shown that the Casimir-Polder (CP) force between two uniformly accelerating atoms in their ground state exhibits a cross-over from a short-distance thermal behavior to a long-distance non thermal behavior, with respect to a reference length identified with $z_a=\frac{c^2}{a}$, where $a$ is the proper acceleration of the atoms. This characteristic length scale coincides with the breakdown of a local approximate description of the two-body system in terms of a Minkowskian space-time \cite{MNP14}.
Indeed, Casimir-Polder forces between neutral atoms arise from the retarded interaction among the dipoles induced and correlated by the zero-point  field fluctuations, and the field quanta mediating the interaction between the two atoms are affected by the non-inertial character of relativistic acceleration, at large distances  \cite{NP13, MNP14}.

In this paper we investigate the resonance interaction between two uniformly accelerated identical atoms, one excited and the other in the ground state, prepared in a correlated state (symmetric or antisymmetric) and show that it exhibits a pure non-thermal behavior, carrying no signature of Unruh thermal fluctuations on the interatomic force. Nevertheless, the relativistic acceleration still causes a qualitative change of the distance-dependence of the interaction between the two atoms.

Resonance interactions between atoms occur when one or more atoms are in their excited state and an exchange of real photons is involved \cite{CT98, Salam10}.
If the two atoms are prepared in a factorized state, resonance Casimir-Polder interactions require a fourth-order perturbation theory. In this case, the interaction scales as $R^{-2}$ for large interatomic separations, $R\gg\lambda$ ($R$ being the interatomic distance and $\lambda$ the main wavelength associated to the atomic transitions). These effects have been recently discussed in the literature, in particular in connection with the relevant problem  of the Casimir-Polder interaction between two nonidentical atoms when one of them is in an excited state \cite{Berman15,DGL15,BPRB15,MR15}.
On the other hand, resonance interactions can also occur when two identical atoms are prepared in a correlated (symmetric or antisymmetric) state, and in this case
they manifest as a second-order effect in the electric charge.
Such interactions scale as $R^{-1}$ in the far zone, and thus they are of a longer range compared with dispersion interactions.
Recently, the possibility to enhance resonance forces between atoms placed
in nano-structured materials such as a photonic crystal has been discussed \cite{IFTPRP14}.
Also, such effects have been investigated in relation to the resonant energy transfer between molecules, and it has been argued that they could play a fundamental role in some biological coherent processes \cite{JA00, PP13, RLBBLB14}.

As mentioned, in this paper we shall consider two identical atoms prepared in a correlated (symmetric or antisymmetric) state and uniformly accelerating in vacuum, and investigate the resonance interaction between them. The radiative properties of uniformly accelerated atoms prepared in a maximally entangled state, have been investigated in the literature until very recently \cite{HY15, MS15}, but here their
resonance interaction is discussed.  A motivation of our work is to explore situations where the effects of the acceleration could be exclusively non-thermal and manifest sharp differences with respect to Unruh thermal-like effects. 
Since the atoms are prepared in a correlated {\em Bell-type} state, our calculation requires only second-order perturbation theory.  Also, since resonance interactions are much more intense than dispersion Casimir-Polder interactions, they could be a suitable candidate for observing the effect of accelerated motion in quantum field theory.

We first consider the atoms interacting with the scalar field and then we generalize our investigation to the case of the electromagnetic field.
We show that specific features, not present for inertial atoms, appear in the resonance interaction as a consequence of the acceleration; specifically, a different scaling of the interaction energy with the distance and a dependence on the acceleration, when compared to the \textquotedblleft Unruh-thermal" Casimir-Polder interaction case \cite{MNP14}.
Following the procedure adopted in \cite{DDC82,DDC84, AM94, AM95}, we separate, at second order in perturbation theory, the contributions of vacuum fluctuations and radiation reaction to the resonance interaction energy between the two atoms. We show, both for the scalar and electromagnetic cases, that the
resonance interaction is related only to the radiation-reaction contribution.
Thus, Unruh thermal fluctuations do not affect the resonance interatomic interaction. Also, we show that in the limit of large accelerations, the distance dependence of the resonance interaction  changes qualitatively.
Thus our results permit one to highlight non-thermal signatures of the atomic acceleration through the second-order resonance interaction between identical atoms.

The paper is organized as follows. In Sec. \ref{sec:2}  we introduce our model and discuss the resonance interaction between two  correlated
accelerated atoms interacting with the scalar field in the vacuum state. In Sec. \ref{sec:3} we generalize our procedure to the more realistic case of accelerated atoms
interacting with the electromagnetic field. Finally, Sec. \ref{sec:4} is devoted to our conclusions and perspectives. Details of some calculations are given in the Appendix.\\

\section{\label{sec:2}Resonance interaction energy between accelerated atoms: the scalar field case}

Let us consider two identical atoms (labelled as $A$ and $B$) modeled as point-like systems
with two internal energy levels, $\mp \frac 1 2\hbar \w0$, associated with the eigenstates  $\g$ and $\e$,  respectively, and separated by a distance $z$.
We assume that the two atoms are accelerating with the same uniform acceleration along two parallel trajectories, $x_A(\tau)$ and $x_B(\tau)$ (therefore their distance is constant), and interacting locally with a real massless scalar field in its vacuum state.
Also, as usual, we suppose that $\w0$ includes any
direct modification of the atomic transition frequency due to the
accelerated motion.
The Hamiltonian describing the
atom-field interacting system in the instantaneous inertial frame of
the two atoms is ($\tau$ is a proper time) \cite{AM94, AM95}
\begin{eqnarray}
&\ &H(\tau)=\hbar\w0\sigma_{3}^{A}(\tau)+\hbar\w0\sigma_{3}^{B}(\tau)
+\sum_{\bk} \hbar \wk\akd\ak
\frac{dt}{d\tau}\nonumber\\
&\ &
+ \lambda(\sigma_{2}^{A}(\tau)\phi(x_{A}(\tau)) +\sigma_{2}^{B}(\tau)\phi(x_{B}(\tau)) ) \, ,
\label{eq:1}
\end{eqnarray}
where $\sigma_{i}^{A(B)}\,\, (i=1,2,3)$ are pseudospin operators in the Hilbert space of the internal degrees of freedom of atoms $A$ and $B$, and $\akd$, $\ak$ are the creation and annihilation bosonic operators of the scalar field
\begin{equation}
\phi({\bf x},t)= \sum_{\bk} \sqrt\frac{\hbar}{2V\wk}\left [\ak(t)e^{i\bk\cdot\bx} + \akd(t)e^{-i\bk\cdot\bx}\right].
\label{eq:2}
\end{equation}

We want to calculate the resonance energy shift of the system of the two accelerated atoms, due to their interaction with the scalar field.
As it is known, the interaction between atoms can be interpreted in terms of radiation source fields, as well as of vacuum field fluctuations \cite{Milonni}. In this paper, we
obtain the resonance interaction following a procedure proposed in \cite{DDC82, DDC84, AM94, AM95}, which consists of separating at a given order in perturbation theory, the contributions of {\emph{vacuum fluctuations}} ({\em vf}) and  {\emph{self-reaction}} ({\em sr}) to the time evolution of a generic atomic observable.
The method consists of rewriting these two different
contributions in terms of two effective Hamiltonians,
$H^{eff}_{vf}$ and $H^{eff}_{sr}$, and then computing their contribution to
the resonant energy shift.
This approach has been used to investigate the effect of the atomic acceleration on the radiative properties of single atoms  \cite{YZ06, Passante98, Rizzuto07, ZY10, MNPRS14, RS09} and has been recently generalized to calculate the Casimir-Polder interaction between two uniformly accelerated atoms \cite{MNP14}.

We first derive the effective Hamiltonians $H^{eff}_{vf}$ and $H^{eff}_{sr}$ at second order in the atom-field coupling.
After some algebra (details are given in the Appendix), we find the following expressions
\begin{widetext}
\begin{equation}
\Bigl(H_A^{eff}\Bigr)_{vf}=-i\frac{\lambda^2}{2\hbar}\int_{\tau_0}^{\tau}d\tau' C^F(\xAt,\xAtp)
\Bigl[\sigma^{f}_{2,A}(\tau),\sigma^{f}_{2,A}(\tau')\Bigr],
\label{eq:5}
\end{equation}
and
\begin{eqnarray}
&\ &\Bigl(H_A^{eff}\Bigr)_{sr}=-i\frac{\lambda^2}{2\hbar}\Biggl\}\int_{\tau_0}^{\tau}d\tau' \chi^F(\xAt,\xAtp)
\Bigl\{\sigma^{f}_{2,A}(\tau),\sigma^{f}_{2,A}(\tau')\Bigr\}+\int_{\tau_0}^{\tau}d\tau'\chi^F(\xAt,\xBtp)\Bigl\{\sigma^{f}_{2,A}(\tau),\sigma^{f}_{2,B}(\tau')\Bigr\}\Biggr\}
\label{eq:6} \, ,
\end{eqnarray}
\end{widetext}
where we have introduced the statistical functions for the scalar field given in the Appendix. Analogous expressions for $(H_B^{eff})_{vf/sr}$ are obtained by exchange of $A$ and $B$ in the relations above. 

The resonance interaction between two atoms moving on the stationary trajectories $x_A(\tau)$ and $x_B(\tau)$, is then obtained by evaluating the expectation value of the effective Hamiltonians $H^{eff}_{vf/sr}=(H_A^{eff})_{vf/sr}+(H_B^{eff})_{vf/sr}$, on the correlated state of the two atoms, and taking into account only the terms depending on the atomic separation.

In order to do that, we suppose the two atoms prepared in one of the correlated states
\begin{equation}
\mid \psi_{\pm} \rangle = \frac 1{\sqrt{2}} \left( \mid g_A, e_B \rangle \pm \mid e_A, g_B\rangle \right) \, ,
\label{eq:9}
\end{equation}
where, as mentioned before, $g$ $(e)$ indicates the ground (excited) state of the atom.
In these states the atomic excitation is delocalized among the two atoms. The symmetrical state is called a super-radiant state because in the Dicke model its decay rate is larger than that of the individual atoms, yielding a collective spontaneous decay \cite{Dicke}. On the contrary, the antisymmetric superposition is called a subradiant state because its spontaneous decay is inhibited.
These two correlated states are degenerate, and thus perturbation theory for degenerate states should be used, in principle. However, it is possible to show
that if the dipole matrix elements of the two identical atoms are equal in modulus and the photon dispersion relation is symmetric between $\bk$ and $-\bk$, the symmetric and antisymmetric subspaces do not mix each other at the order considered. Then, in this case the degenerate symmetric and antisymmetric states can be treated independently and the second-order energy shift is the same as in the nondegenerate perturbation theory for both states.
Different methods to obtain super- and subradiant states for two-level systems, have been proposed (see, for example, \cite{LSLFWB13, LFLSBW13}).
On the other hand, it has been recently shown that entanglement between accelerated systems can be induced by the Unruh bath \cite{HY15, BF04}. To obtain the resonance energy shift for the system considered, we now evaluate the expectation values of
$H^{{eff}}_{vf}$ and $H^{{eff}}_{sr}$ on state  (\ref{eq:9}) (symmetrization with respect to $A\leftrightarrows B$ is necessary to obtain the total energy shift).  After some algebra, we get
\begin{widetext}
\begin{eqnarray}
&\ &(\delta E)_{vf}= (\delta E_A)_{vf}+(\delta E_B)_{vf}=-\frac{i\lambda^2}{\hbar}\int_{\tau_0}^{\tau}d\tau'C^F(\xAt,\xAtp)\chi_{A}(\tau,\tau')
+(A \leftrightarrows B \,\mbox{terms}) \, ,
\label{eq:3}\\
&\ &(\delta E)_{sr}= (\delta E_A)_{sr}+(\delta E_B)_{sr}=-\frac{i\lambda^2}{\hbar}\int_{\tau_0}^{\tau}d\tau'\chi^F(\xAt,\xAtp)C_{A}(\tau,\tau')\nonumber\\
&\ &-\frac{i\lambda^2}{\hbar}\int_{\tau_0}^{\tau}d\tau'\chi^F(\xAt,\xBtp)C_{A,B}(\tau,\tau')
+(A \leftrightarrows B \,\mbox{terms}) \, ,
\label{eq:10}
\end{eqnarray}
\end{widetext}
where we have introduced the atomic statistical functions
\begin{eqnarray}
C_{A,B}(\tau,\tau')=\frac 1 2 \langle \psi_{\pm}\vert\Bigl\{\sigma^{f}_{2,A}(\tau),\sigma^{f}_{2,B}(\tau')\Bigr\}\vert\psi_{\pm}\rangle \, , \label{eq:11} \\
\chi_{A,B}(\tau,\tau')=\frac 1 2 \langle \psi_{\pm}\vert\Bigl [\sigma^{f}_{2,A}(\tau),\sigma^{f}_{2,B}(\tau')\Bigr]\vert\psi_{\pm}\rangle \, .
\label{eq:11a}
\end{eqnarray}

The vacuum fluctuation contribution (\ref{eq:3}) gives only the Lamb shift of each atom, as if the other atom were absent, and thus it does not contribute to the interatomic interaction energy.
Similarly, the first term in (\ref{eq:10}) describes the self-reaction contribution to the Lamb shift of each atom. On the contrary, the second term in (\ref{eq:10}) is the relevant one for the resonance interaction; it describes the interaction of one atom with its own emitted field as modified by the presence of the other atom (as expressed by $\chi^F(\xAt,\xBtp)$). It depends on the distance between the two atoms, and it is the only contribution to the interatomic interaction at the order considered.

The result above shows that the resonance interaction is due to the radiation reaction contribution $\delta E_{sr}$ only. This is indeed expected from a physical ground. The resonance interaction originates from the exchange of one photon between the two atoms that are in a correlated state; thus, contrary to the dispersion interaction where the atomic correlation is induced by vacuum fluctuations, it is not related to nonlocal vacuum field correlations, but is entirely due to the field radiated by the atoms (source field). This peculiarity has relevant consequences when we consider the resonance interaction between accelerated atoms; in fact, we will now show that resonance interactions do not display any signature of the {\em Unruh thermal} effect (which is related to the correlations of vacuum field fluctuations in the locally inertial frame). Nevertheless, the atomic acceleration will cause a qualitative change of the interaction between the two atoms.
This situation should be compared with the case of the Casimir-Polder dispersion interaction between two uncorrelated atoms, where atomic dipoles are induced and correlated by the zero-point fluctuations of the field and the interaction is directly related with the spatial correlations of vacuum fluctuations.

The procedure outlined above is general and valid for any arbitrary stationary trajectory.
We now focus on the case of two atoms moving with the same uniform acceleration along the trajectory

\begin{eqnarray}
&\ &t(\tau)=\frac{c}{a}\sinh\frac{a\tau}{c},\,\,
x_{A/B}(\tau)=\frac{c^2}{a}\cosh\frac{a\tau}{c},\nonumber\\
&\ & y_{A/B}(\tau)=0\,\,\,z_A(\tau)=z_A\,\,\,,z_B(\tau)=z_B \label{eq:17},
\end{eqnarray}

We first evaluate the linear susceptibility of the scalar  field  (\ref{A12}) and the atomic correlation function (\ref{eq:11}). We have
\begin{widetext}
\begin{eqnarray}
&\ &\chi^F(\xAt,\xBtp)=-\frac{\hbar}{8\pi^2 c^2}\frac{1}{z\sqrt{\emph{N}(z,a)}}\int_0^{\infty}d\omega g(\omega,z,a)
(e^{i\omega (\tau-\tau')}-e^{-i\omega (\tau-\tau')}),
\label{eq:18}
\end{eqnarray}
\end{widetext}
and
\begin{eqnarray}
&\ &C_{A,B}(\tau',\tau) = \pm\frac{1}{ 8} (e^{i\omega_0 (\tau-\tau')}+e^{-i\omega_0 (\tau-\tau')}) \, ,
\label{eq:14}
\end{eqnarray}
where we have defined $\emph{N}(z,a)=1+(z a/2c^2)^2$ and $g(\omega,z,a)=\sin(\frac{2\omega c}{a} \sinh^{-1}(\frac{z a}{2c^2}))$. Finally, $z=z_B-z_A$.
The upper (lower) sign in (\ref{eq:14}) refers to the symmetric (antisymmetric) superposition in (\ref{eq:9}), respectively.

Now, substituting (\ref{eq:18}) and (\ref{eq:14}) in Eq. (\ref{eq:10}) and taking into account only $z-$dependent terms, we finally obtain the resonance interaction between the two accelerated atoms
\begin{widetext}
\begin{eqnarray}
&\ &\delta E =\left (\delta E_{A}\right)_{sr}+\left (\delta E_{B}\right)_{sr} =\mp\frac{\lambda^2}{16\pi^2 c^2}\frac{1}{z\sqrt{\emph{N}(z,a)}}\int_0^{\infty}d\omega g(\omega,z,a)\Bigl(\frac{1}{\omega+\omega_0}+\frac{1}{\omega-\omega_0} \Bigr) \, .
\label{eq:19}
\end{eqnarray}
\end{widetext}
We have also taken the limits $\tau_0\rightarrow -\infty$ and $\tau\rightarrow\infty$ in the integral above, since we are considering the atoms uniformly accelerating in vacuum for all times.

The integral in the equation above can be evaluated exactly, giving
\begin{eqnarray}
&\ &\delta E = \mp\frac{\lambda^2}{16\pi c^2}\frac{1}{z\sqrt{N(z,a)}}\cos\Bigl(\frac{2\omega_0 c}{a} \sinh^{-1}(\frac{z a}{2c^2})\Bigr)
\label{eq:20}
\end{eqnarray}
Equation (\ref{eq:20}) is the main result of this section.
Since the interaction energy is entirely due to the radiation reaction contribution, the atomic acceleration does not yield thermal signatures in the resonance interaction; its only effect is enclosed in the normalization factor $N(z,a)$ and in the function $g(\omega,z,a)$.
More precisely, a comparison with the Casimir-Polder interaction between two accelerated atoms as in \cite{MNP14} shows that while in that case the effect of acceleration is a {\em thermal} correction with the Unruh temperature $T=\hbar a/2\pi ck_B $ (due to the presence of a factor $\coth(\pi c\omega/a)$ in the vacuum fluctuations contribution), in the present case (resonance interaction) the effects of atomic acceleration are not in the form of such a {\em thermal term}.
Most importantly, the presence of the factor $N(z,a)$ yields a change of the scaling of the interaction with the distance. Indeed, we can identify a characteristic length scale, $z_a=c^2/a$,  in analogy to the result of \cite{MNP14} for the dispersion interaction. For distances smaller than $z_a$, it is possible to find a local inertial frame where the linear susceptibility of field is fairly well described by its static counterpart; on the other hand, signals spreading over distances larger than $z_a$ may be affected by the non-inertial character of relativistic acceleration. Accordingly, we expect that relativistic accelerations can deeply modify the qualitative behavior of the resonance-interaction energy,
in particular its distance dependence.
In fact, in the limit $z\gg c^2/a$, we get
\begin{eqnarray}
&\ &\delta E \simeq \mp\frac{\lambda^2}{8\pi}\frac{1}{z^2 a }\cos\Bigl(\frac{2\omega_0 c}{a} \ln\left(\frac{z a}{c^2} \right)\Bigr),
\label{eq:21}
\end{eqnarray}
while for $z\ll c^2/a$, we recover the inertial result
\begin{eqnarray}
&\ &\delta E \simeq \mp\frac{\lambda^2}{16\pi c^2}\frac{1}{z}\cos\Bigl(\frac{\omega_0 z}{c}\Bigr).
\label{eq:22}
\end{eqnarray}

Thus, the resonance interaction strongly bears signatures of a relativistic acceleration, resulting in a different power-law distance dependence, scaling at large distances as $z^{-2}$ rather than the usual $z^{-1}$ of the inertial case (\ref{eq:22}).
This result should be compared with that obtained in \cite{MNP14}, where it was shown that, as a consequence of metric effects, the scalar Casimir-Polder interaction between two uniformly accelerated ground-state atoms, was characterized by a $z^{-4}$ power law decay,  for distances $z\gg c^2/a$.

We also stress that Eq. (\ref{eq:21}) exhibits a global overall pre-factor depending on the inverse of the acceleration, while the  \textquotedblleft thermal-Unruh" analogy would have suggested the presence of a Unruh term with temperature $T_U=a/2\pi$, directly proportional to the acceleration \cite{MNP14}.
Thus, our result shows that it is possible to single out metric effects associated to relativistic accelerations from the usual \textquotedblleft Unruh thermal-like" effects.

The limit $z\ll z_a=c^2/a$ in (\ref{eq:22}) gives back the resonance interaction for the inertial case. For typical interatomic distances $z\sim 10^{-6}$m, this limit is valid also for considerable accelerations, thus suggesting that the resonance interaction is almost insensitive to the atomic acceleration when $z\ll z_a$.
Actually, such a behavior can be expected from the following physical considerations. Resonance interactions arise from the exchange of real photons between the atoms.
If the distance between the atoms is much smaller than $z_a$, during the time taken by the photon emitted by one atom to reach the other atom ($t\sim z/c$), the accelerating atoms move of a distance $x$ which is much smaller than their interatomic distance $z$; thus the photon mediating the interaction cannot discern the atomic motion, and the interaction is essentially the same between stationary atoms.

Finally, it is worth noting from (\ref{eq:22}) that the inertial resonance interaction decreases as $z^{-1}$ for any interatomic distance. This is  a peculiarity of the scalar field model we have considered. The distance behavior is different in the case of the electromagnetic field, where we can distinguish a near and a far zone, as we shall discuss in the next section.

\section{\label{sec:3} The electromagnetic field case}

We now extend our investigation to the case of two uniformly accelerated atoms interacting with the electromagnetic field in the vacuum state.

To describe our system, we adopt the Hamiltonian in the multipolar coupling scheme and in the dipole approximation
\begin{eqnarray}
&\ & H= H_{A}+H_{B}+\sum_{{\bf k}j}\hbar \omega_k a^{\dagger}_{{\bf k}j}a_{{\bf k}j}\frac{dt}{d\tau} \nonumber\\
&\  & - \boldsymbol\mu_{A}(\tau)\cdot{\bf E}(x_{A}(\tau))-\boldsymbol\mu_{B}(\tau)\cdot{\bf E}(x_{B}(\tau))\label{eq:11ab} \, ,
\label{eq:35}
\end{eqnarray}
where j=1,2 is the polarization index,
${\bf E}(x(\tau))$ is the electric field operator, and $\boldsymbol\mu=e{\bf r}$ is the atomic dipole moment operator. As in the previous section, the atoms, with transition frequency
$\omega_0$, have a uniform acceleration $a$ along the $x$ direction and are separated by a constant distance $z$ along the $z$ direction, while $x_{A/B}(\tau )$ are the trajectories of the two atoms.

As discussed before, the resonant interaction energy is related only to the radiation reaction contribution and is obtained from the expectation value of the effective Hamiltonian $(H^{eff}_A)_{sr}+(H^{eff}_B)_{sr}$
on the state $\vert \psi_{\pm}\rangle$
\begin{eqnarray}
&\ &\delta E = -\frac{e^2}{2}\int_{\tau_0}^{\tau}d\tau' \chi^F_{\ell m }(x_A(\tau),x_B(\tau')) C^{A/B}_{\ell m}(\tau,\tau') \nonumber\\
&\ &+ (A \leftrightarrows B \,\mbox{terms}) \, .
\label{eq:36}
\end{eqnarray}
In order to evaluate this quantity, we first obtain the statistical functions for the field and the atoms.
The susceptibility of the electromagnetic field in the accelerated frame can be obtained from
the two-point field correlation function in the proper reference frame of the two accelerated atoms (Rindler noise) \cite{Takagi}. After lengthy calculations involving Lorentz transformations of the electromagnetic field, we obtain

\begin{eqnarray}
&\ &\mathcal{G}_{\ell m}(x_A(\tau),x_B(\tau'))=\langle 0|E_{\ell}(x_A(\tau)) E_m(x_B(\tau'))|0\rangle \nonumber\\
&\ &= \frac{\hbar a^4}{4\pi c^7}
\frac{1}{(\sinh^2 \frac{a(\tau-\tau'-i\epsilon)}{2c}-(\frac{z a}{2c^2})^2)^3}\nonumber\\
&\ &\times \Biggl\{[\delta_{\ell m}-\frac{za}{c^2}(n_m q_{\ell} - n_{\ell}q_m)]\sinh^2\frac{a(\tau-\tau')}{2c}\nonumber\\
&\ &+(\frac{za}{2c^2})^2[\delta_{\ell m}-2n_{\ell}n_m]\nonumber\\
&\ &\times\biggr[1+2(\delta_{\ell m}-q_{\ell}q_m)\sinh^2 \frac{a(\tau-\tau')}{2c}\Biggr]
\Biggr\}
\label{eq:71}
\end{eqnarray}
($\ell,m=x,y,z$). ${\bf n}=(0,0,1)$ is the unit vector along the $z$ direction and ${\bf q}=(1,0,0)$ is the unit vector along the direction of acceleration, $x$.
A simple calculation shows that the only nonzero components of $\mathcal{G}_{\ell m}$ are the $xx$, $yy$, $zz$, $xz$, and $zx$ components.
In particular, $\mathcal{G}_{\ell\ell}(x_A(\tau),x_B(\tau'))\neq \mathcal{G}_{mm}(x_A(\tau),x_B(\tau'))$ (for $\ell\neq m$);
therefore, the Rindler noise
evaluated on the atomic trajectories of the two accelerated atoms, is not isotropic and displays a non-diagonal component.  A similar anisotropy is not present in the case of a single uniformly accelerated atom in the unbounded space, where it is possible to show that
the Rindler noise is isotropic.
Actually, in our system we have two characteristic spatial directions, namely the direction of the acceleration and the distance between the atoms;
in this sense, the anisotropic aspect of the Rindler function can be ascribed to the spatially extended character of the two-particle system considered.

From Eq. (\ref{eq:71}), we can obtain the linear susceptibility of the electromagnetic field in the proper reference frame,

\begin{eqnarray}
&\ &\chi_{\ell m}^F(x_A(\tau),x_B(\tau')) =\frac{i}{\hbar}\langle 0|[E_{\ell}(x_A(\tau)),E_m(x_B(\tau'))]|0\rangle\, .
\end{eqnarray}
\\
Its expression, in the form of an integral over frequencies, is

\begin{eqnarray}
&\ &\chi_{\ell m}(x_A(\tau),x_B(\tau)) =\frac{1}{2\pi z^3}\int_0^{\infty}d\omega(e^{i\omega u}-e^{-i\omega u})\nonumber\\
&\ &\times\Biggl[\biggl(f_{\ell m}(a,z,\omega)+\frac{az}{2c^2}F_{\ell m}(a,z,\omega)\biggr)\cos(\omega S)\Biggr.\nonumber\\
&\ &\Biggl.+\biggl(g_{\ell m}(a,z,\omega)+\frac{az}{2c^2}G_{\ell,m}(a,z,\omega)\biggr)\sin(\omega S)\Biggr]\, ,
\label{eq:70}
\end{eqnarray}
\\
\noindent where  $u=\tau-\tp$ and
$ S=\frac{2c}{a}\sinh^{-1}\left(\frac{z a}{2c^2} \right)$. We have also defined the functions

\begin{widetext}
\begin{eqnarray}
f_{\ell m}(a,z,\omega)&=&\frac{1}{N^2}\frac{\omega z}{c} \Biggl[(\delta_{\ell m}-3n_{\ell}n_m)+\frac{a^2z^2}{4c^4}\biggl(2(\delta_{\ell m}+q_{\ell}q_m-n_{\ell}n_m)+(\delta_{\ell m}-q_{\ell}q_m-2n_{\ell}n_m)\left(1+\frac{a^2z^2}{2c^4}\right)\biggr)\Biggr]\;,
\label{eq:70a}
\end{eqnarray}
\begin{eqnarray}
&\ &g_{\ell m}(a, z,\omega)=-\frac{1}{N^{5/2}}\biggl[\delta_{\ell m}\biggl(1+\frac{a^2 z^2}{4c^4}\biggr) + q_{\ell}q_m  \frac{a^2 z^2}{4 c^4}\biggl(1+\frac{a^2z^2}{c^4}\biggr) - 3n_{\ell}n_m \biggl(1+\frac{a^2z^2}{2c^4}\biggr)\biggr]\nonumber\\
&\ & + \frac{\omega^2 z^2}{c^2N^{3/2}}\biggl[\delta_{\ell m}\biggl(1+\frac{a^2 z^2}{4c^4}\biggr) - q_{\ell}q_m \frac{a^2 z^2}{4 c^4} - n_{\ell}n_m \biggl(1+\frac{a^2 z^2}{2c^4}\biggr)\biggr]\, ,
\label{eq:70b}
\end{eqnarray}
\end{widetext}
\begin{eqnarray}
\label{eq:70c}
&\ &F_{\ell m}(a, z,\omega)=\biggl(n_m q_{\ell}-n_{\ell}q_m\biggr)\frac{\omega z}{cN^2} \biggl(1-\frac{a^2 z^2}{2c^4}\biggr)\, , \\
&\ &G_{\ell m}(a, z,\omega)=\biggl(n_m q_{\ell}-n_{\ell}q_m\biggr) \frac{1}{N^{5/2}}\nonumber\\
&\ &\times \biggl(1+\frac{a^2z^2}{c^4}+\frac{\omega^2z^2}{c^2}\left(1+\frac{a^2z^2}{4c^4}\right)\biggr)\, .
\label{eq:70d}
\end{eqnarray}

The functions $f_{\ell m}(a, z,\omega)$ and $g_{\ell m}(a, z,\omega)$ are non-vanishing only for $\ell=m$; on the contrary,  the functions $F_{\ell m}(a, z,\omega)$ and $G_{\ell m}(a, z,\omega)$ have nonvanishing non-diagonal terms, yielding non-diagonal contributions to the field susceptivity.

The symmetric correlation function for the  atoms is
\begin{eqnarray}
&\ &C^{A/B}_{\ell m}(\tau,\tau') = \frac{1}{2}\langle \psi_{\pm}\vert \left\{r_{\ell}^A(\tau),r_m^B(\tau') \right\}\vert\psi_{\pm}\rangle\nonumber\\
&\ &=\pm \frac{1}{2}\left(e^{i\omega_0(\tau-\tau')}+e^{-i\omega_0(\tau-\tau')}\right) (r^A_{ge})_{\ell}(r^B_{eg})_{m} \, ,
\label{eq:72}
\end{eqnarray}
where subscripts $eg$ indicate a matrix element between the atomic excited and ground states.

The resonance interaction between the two accelerated atoms is now obtained substituting Eqs. (\ref{eq:70}) and (\ref{eq:72}) into Eq. (\ref{eq:36}), and taking the limits $\tau_0\rightarrow -\infty$, $\tau\rightarrow\infty$. After simple algebraic manipulations,  we obtain

\begin{widetext}
\begin{eqnarray}
\delta E =\pm(\mu^A_{eg})_{\ell}(\mu^B_{ge})_m V_{\ell m}(a, z,\omega_0) \pm(\mu^A_{eg})_{\ell}(\mu^B_{ge})_m W_{\ell m}(a, z,\omega_0)\, ,
\label{eq:24}
\end{eqnarray}
\end{widetext}
where the explicit expressions of $V_{\ell m}(a, z,\omega_0)$ and $W_{\ell m}(a, z,\omega_0) $ are

\begin{eqnarray}
&\ & V_{\ell m}(a, z,\omega_0)=\frac{1}{z^3}\biggl[f_{\ell m}(a,z,\omega_0)\sin(\omega_0 S)\biggr.\nonumber\\
&\ &\biggl. -g_{\ell m}(a,z,\omega_0)\cos(\omega_0 S)\biggr] \, ,
\end{eqnarray}

\begin{eqnarray}
&\ & W_{\ell m}(a, z,\omega_0)=\frac{a}{2z^2c^2}\biggl[F_{\ell m}(a,z,\omega_0)\sin(\omega_0 S)\biggr.\nonumber\\
&\ & \biggl.-G_{\ell m}(a,z,\omega_0)\cos(\omega_0 S)\biggr]\, .
\end{eqnarray}

These quantities explicitly depend on the atomic acceleration and are the generalization of the stationary interaction potential to the case of accelerated atoms.

The expression given above is valid for
any value of $az/c^2$.  As in the scalar field case, we now investigate two limiting cases, $z\ll c^2/a$ and $z\gg c^2/a$.
It is easy to show that, for $za/c^2\ll 1$, the linear susceptibility (\ref{eq:70}) is fairly well described by its stationary counterpart. Therefore, at the lowest order in $za/c^2$, we recover the known inertial resonance interaction \cite{CT98}
\begin{eqnarray}
&\ &\delta E =\pm(\mu^A_{eg})_{\ell}(\mu^B_{ge})_m V_{\ell m}(\omega_{0},z) \, ,
\label{eq:30}
\end{eqnarray}
where
$V_{\ell m}(\omega_{0},z)$
is the well-known tensor potential
\begin{eqnarray}
&\ &V_{\ell m}=\frac{1}{z^3}\Biggl\{\left(\delta_{\ell m}-3n_{\ell}n_m\right)\biggl[\cos(\omega_0 z/c)+\frac{z\omega_0}{c}\nonumber\\
&\ &\times\sin(\omega_0 z/c)\biggr]- (\delta_{\ell m}-n_{\ell} n_m)\frac{z^2\omega_0^2}{c^2}\cos(\omega_0 z/c)\Biggr\} \, ,
\end{eqnarray}

In particular, Eq. (\ref{eq:30}) yields a potential energy as  $\sim z^{-1}$ in the far-zone $R\gg \lambda$, and as $\sim z^{-3}$  in the near-zone $R\ll\lambda$,
with $\lambda$ being the atomic transition wavelength \cite{CT98}.

On the other hand, at higher orders in $az/c^2$,  the corrections due to the atomic accelerated motion give a qualitative change of the distance dependence of the resonance interaction, scaling with a different power law, as expected.
For example, in the case of atomic dipoles with the same direction along one of the axies $x,y,z$, for $za/c^2 \gg 1$ we obtain from (\ref{eq:24})
[in this case only the term containing $V_{\ell m}(a, z,\omega_0)$ contributes],

\begin{eqnarray}
&\ &\delta E \simeq \pm (\mu_{eg}^{A})_{\ell}(\mu_{ge}^{B})_m \frac{1}{z^3}\Biggl\{(\delta_{\ell m}-q_{\ell}q_m-2n_{\ell}n_m)\Biggr.\nonumber\\
&\ &\Biggl.\times\biggl[\frac{2\omega_0z}{c} \sin\Bigl(\frac{2\omega_0 c}{a}\ln\biggl(\frac{az}{c^2}\biggr)\Bigl)-\frac{\omega_0^2 z^2}{c^2}\left(\frac{2c^2}{z a}\right)\biggr.\Biggr.
\nonumber\\
&\ &\Biggl.\biggl.\times\cos\Bigl(\frac{2\omega_0 c}{a}\ln\left(\frac{az}{c^2}\right)\Bigr)\biggr]+q_{\ell}q_m\left(\frac{8c^2}{a z}\right)\Biggr.\nonumber\\
&\ &\Biggl.\times\cos\Bigl(\frac{2\omega_0 c}{a}\ln\biggl(\frac{az}{c^2}\biggr)\Bigr)\Biggr\}\, .
\label{eq:26}
\end{eqnarray}

Our result (\ref{eq:26}) shows that the far-zone resonance interaction between two accelerated atoms decreases with the distance as $z^{-2}$ if the two dipoles are along $z$ or $y$ or as $z^{-4}$ it they are along $x$,
compared with the $z^{-1}$ dependence for atoms at rest.
Also, a comparison with the case of the scalar field discussed in the previous section, shows the emergence of another change of the resonance-interaction behavior
originating from the structure of the field susceptibility, ultimately related to the vector nature of the electromagnetic field and to the spatially extended character of our two-atom system (and not only to the presence of the metric factor $N(z,a)$).
In fact, Eq. (\ref{eq:26}) clearly shows that it is possible to control the effects of atomic acceleration on the resonance interaction by an appropriate choice of the orientation of the two dipole moments; for example, the resonance interaction is strongly suppressed if the two dipole moments are oriented along the $x$ direction, which is along the direction of acceleration.
Also, when the dipoles are orthogonal to each other, with one of them along $z$ and the other one in the $(x,y)$ plane, the term in (\ref{eq:24}) containing $V_{\ell m}(a, z,\omega_0)$ vanishes and only that with $W_{\ell m}(a, z,\omega_0)$ survives. Such a  non-diagonal term is present only for $a \neq 0$, and thus its contribution is a unique signature of the accelerated motion yielding, in this specific configuration (where the interaction for stationary atoms is zero), a nonvanishing interaction energy.
These properties suggest further possibilities  to single out the effect of the acceleration by an appropriate choice of the orientation of the dipole moments of the two accelerated atoms, which are not present for a single accelerated atom in the unbounded space.

\section{\label{sec:4} Conclusions and Future Perspectives.} In this paper we have investigated the resonance interaction between two uniformly accelerated atoms, one excited and the other in the ground state, prepared in a correlated (symmetrical or anti-symmetrical) state, and interacting with the massless scalar field or the electromagnetic field in their vacuum state. We have considered the contributions of vacuum fluctuations and of the radiation reaction field to the resonance interaction, and shown that the Unruh thermal fluctuations do not affect the interatomic interaction, which is exclusively given by the radiation reaction term. We have shown that non-thermal effects appear and that the acceleration yields a change of the distance dependence of the resonance interaction. Specifically, we have shown that these non-thermal effects, related to the non-inertial character of acceleration, result in a different scaling with the distance and a different dependence on the acceleration as expected from the known Unruh-temperature equivalence.

Our results open the way for new future developments, for example the impact of the change of interatomic potential on the Dicke phase transition or on macroscopic phenomena such as thermodynamical properties of systems composed of many accelerating particles. These points could be a promising direction to highlight non-thermal signatures of relativistic accelerations in a many-body context.

\section*{ACKNOWLEDGMENTS}
Financial support by the Julian Schwinger Foundation, MIUR, University of Palermo, is gratefully acknowledged.
J. M.  acknowledges support from the Alexander Von Humboldt Foundation. W. Z. acknowledges financial support from China Scholarship Council (CSC)

\appendix
\renewcommand{\theequation}{\thesection.\arabic{equation}}
\section{Vacuum fluctuations and radiation reaction: the general formalism}

In this Appendix, we briefly recall the procedure of Dalibard et al \cite{DDC84}, that we have used in this paper to separate the contributions of vacuum fluctuations and radiation reaction to the resonance interaction between the two atoms. 
We assume a linear coupling between atoms and field, as in Eq. (\ref{eq:1}).

The solution of Heisenberg equations for the field operators can be separated in a {\em free} part  ($\ak^{f}$), which is present even in the absence of interaction, and a {\em source}
part  ($\ak^{s}$), which is related to the interaction between atoms and field,
\begin{eqnarray}
\ak(t(\tau))=\ak^{f}(t(\tau))+\ak^{s}(t(\tau))
\label{A1}
\end{eqnarray}
with
\begin{eqnarray}
&\ &\ak^{f}(t(\tau))=\ak(t(\tau_0))e^{-i\wk(t(\tau)-t(\tau_0))}
\label{A2}
\end{eqnarray}
and
\begin{eqnarray}
&\ & \ak^{s}(t(\tau))=
\frac{i\lambda}{\hbar}\Biggl\{\int_{\tau_0}^{\tau}d\tau'\sigma_{2,A}\bigl[\phi(\xAtp),\ak(t(\tp))\Bigr]\Biggr.\nonumber\\
&\ &\Biggl.\times e^{-i\wk(t(\tau)-t(\tp))}+\int_{\tau_0}^{\tau}d\tau'\sigma_{2,B}\bigl[\phi(\xBtp),\ak(t(\tp))\Bigr]\Biggr.\nonumber\\
&\ &\Biggl.\times e^{-i\wk(t(\tau)-t(\tp))}\Biggr\}
\label{A3}
\end{eqnarray}
(an analogous separation can be performed for the atomic observables).
Then, the equations of motion of a generic atomic observable (pertaining, for example, to the atom $A$), $O_A$, using symmetric ordering between atom and field variables, can be splitted in the {\emph{vacuum fluctuations}} and  {\emph{self-reaction}} contributions
\begin{eqnarray}
&\ &\left(\frac{dO_A}{d\tau}\right)=\left(\frac{dO_A}{d\tau}\right)_{vf}+\left(\frac{dO_A}{d\tau}\right)_{sr},
\label{A4}
\end{eqnarray}
where
\begin{widetext}
\begin{eqnarray}
&\ &\left(\frac{dO_A}{d\tau}\right)_{vf}=\frac{i\lambda}{2\hbar}\Bigl(\phi^{f}(x(\tau))[\sigma_{2,A}(\tau),O_A(\tau)]+[\sigma_{2,A}(\tau),O_A(\tau)]\phi^{f}(x(\tau))\Bigr),
\label{A5}
\end{eqnarray}
and
\begin{eqnarray}
&\ &\left(\frac{dO_A}{d\tau}\right)_{sr}=\frac{i\lambda}{2\hbar}\Bigl(\phi^{s}(x(\tau))[\sigma_{2,A}(\tau),O_A(\tau)]+[\sigma_{2,A}(\tau),O_A(\tau)]\phi^{s}(x(\tau))\Bigr).
\label{A6}
\end{eqnarray}
\end{widetext}
The procedure outlined above is exact.
To obtain  the {\em vf} and {\em sr} contributions to the time evolution of the observable $O_A$ up to second-order in the coupling constant, we now separate the atomic operators $\sigma(\tau)$ and $O_A$ into free and source parts,
and substitute their expressions in the equations above; after some algebra, we get
\begin{widetext}
\begin{eqnarray}
\label{A7}
&\ &\left(\frac{dO_A}{d\tau}\right)_{vf} = \frac{i\lambda}{2\hbar}\Bigl(\phi^{f}(\xAt)[\sigma^{f}_{2,A}(\tau),O^{f}_A(\tau)]+[\sigma^{f}_{2,A}(\tau),O^{f}_A(\tau)]\phi^{f}(\xAt)\Bigr)\nonumber\\
&\ & - \frac{\lambda^2}{2\hbar^2}\int_{\tau_0}^{\tau}d\tau'\bigl\{\phi^{f}(\xAt),\phi^{f}(\xAtp)\Bigr\}[\sigma^{f}_{2,A}(\tau'),[\sigma^{f}_{2,A}(\tau),O^{f}_A(\tau)]]+O(\lambda^3),
\end{eqnarray}
and
\begin{eqnarray}
&\ &\left(\frac{dO_A}{d\tau}\right)_{sr} =
- \frac{\lambda^2}{2\hbar^2}\int_{\tau_0}^{\tau}d\tau'\bigl[\phi^{f}(\xAt),\phi^{f}(\xAtp)\Bigr]\Bigl\{\sigma^{f}_{2,A}(\tau'),[\sigma^{f}_{2,A}(\tau),O^{f}_A(\tau)]\Bigr\}\nonumber\\
&\ &- \frac{\lambda^2}{2\hbar^2}\int_{\tau_0}^{\tau}d\tau'\bigl[\phi^{f}(\xBtp),\phi^{f}(\xAt)\Bigr]\Bigl\{\sigma^{f}_{2,B}(\tp),[\sigma^{f}_{2,A}(\tau),O^{f}_A(\tau)]\Bigr\} +O(\lambda^3).
\label{A8}
\end{eqnarray}
\end{widetext}
The average values of Eqs. (\ref{A7}) and (\ref{A8}) on the vacuum field state $\mid  0 \rangle$, can be expressed in terms of the effective Hamiltonians, $H_{vf}$ and $H_{sr}$,
\begin{widetext}
\begin{equation}
\Bigl(H_A^{eff}\Bigr)_{vf}=-\frac{i\lambda^2}{2\hbar}\int_{\tau_0}^{\tau}d\tau' C^F(\xAt,\xAtp)
\Bigl[\sigma^{f}_{2,A}(\tau),\sigma^{f}_{2,A}(\tp)\Bigr]
\label{A9}
\end{equation}
and
\begin{eqnarray}
&\ &\Bigl(H_A^{eff}\Bigr)_{sr}=-\frac{i\lambda^2}{2\hbar}\int_{\tau_0}^{\tau}d\tp\Biggl\{ \chi^F(\xAt,\xAtp)
\bigl\{\sigma^{f}_{2,A}(\tau),\sigma^{f}_{2,A}(\tp)\bigr\}+\chi^F(\xAt,\xBtp)\bigl\{\sigma^{f}_{2,A}(\tau),\sigma^{f}_{2,B}(\tp)\bigr\}\Biggr\}
\label{A10}\, ,
\end{eqnarray}
\end{widetext}
where we have introduced the statistical functions for the scalar field
\begin{eqnarray}
C^F(\xt,\xtp)= \frac 1 2 \vacd
\{\phi^{\f}(\xt),\phi^{\f}(\xtp)\}\vac \label{A11}\, ,\\
\chi^F(\xt,\xtp)= \frac 1 2 \vacd
[\phi^{\f}(\xt),\phi^{\f}(\xtp)]\vac \, . \label{A12}
\end{eqnarray}
An analogous separation can be performed for the atomic
variables of atom B.
The time evolution of the atomic observables of atom $A$ $(B)$ can be now described by the effective Hamiltonian $H_{A(B)} + (H_{A(B)}^{eff})_{vf} + (H_{A(B)}^{eff})_{sr}$.

\end{document}